\documentstyle[11pt]{article}
\newtheorem{thm}{Theorem}[section]
\newtheorem{cor}[thm]{Corollary} 
\newcommand{\BEQ}{\begin{equation}}
\newcommand{\NEQ}{\end{equation}}
\newcommand{\nn}{\nonumber}

\def\Ebf{{\bf E}\,}

\def\hh{\hat{h}}
\def\kap{{\kappa}}
\def\th{{\theta}}
\def\thh{\hat{\theta}}
\def\part{\partial}
\def\fpth{\part_{f}^p\th}

\def\fqSh{\widehat{\part_{f}^qS}}
\def\fqShk{\widehat{\part_{f}^qS_{\kappa}}}

\def\fqthh{\widehat{\part_{f}^q\th}}
\def\Sh{\hat{S}}
\def\Shb{\bar{\hat{S}}}

\def\nubf{\mathop{\mbox{\boldmath $\nu$}}}

\begin{document}
\bibliographystyle{plain}

\title{Adaptive Smoothing  of the Log-Spectrum \\
with Multiple Tapering 
\footnote{The authors  thank the referees for useful comments. 
Research funded by the U.S. Department of Energy. 
}
}

\author{K.S. Riedel and A. Sidorenko \\
New York University \\
Courant Institute of Mathematical Sciences \\
New York, NY 10012-1185 \\
}
\maketitle

\begin{abstract}
A hybrid estimator of the log-spectral density of a stationary time series
is proposed. First, a multiple taper estimate is performed, followed by
kernel smoothing the log-multiple taper estimate. This procedure reduces the
expected mean square error by $({\pi^2 \over 4})^{4/5}$ over simply
smoothing the log tapered periodogram. 
A data adaptive implementation of a variable bandwidth kernel
smoother is given.
\end{abstract}


\ \\

\section{INTRODUCTION}
\label{INTRO}

We consider a discrete, stationary, Gaussian time
series $\{ x_j , j=1, \ldots N \}$
with a smooth spectral density, $S(f)$, which is bounded away from
zero.
The autocovariance is the Fourier transform of the spectral density:
Cov $[x_j ,x_k ] = \int_{-\frac{1}{2}}^{\frac{1}{2}} S(f)e^{2\pi i(j-k)f} df$.
When the logarithm of the spectral density, $\theta (f) \equiv \ln [S(f)]$,
is desired, two common approaches are: 1) to estimate the spectral
density and then transform to the logarithm; and
2) to smooth the
logarithm of the tapered periodogram. The first approach can be sensitive
to broad-band bias when the  spectral range is large, 
while the second approach inflates the variance of the estimate 
\cite[Ch.~6.15]{PercWald}, \cite{ThomChave}. 
We  propose a combined estimator
of the log-spectral density with the robustness properties of the second
estimator without its variance inflation.

In Section \ref{MTstat}, 
we consider quadratic estimates of the spectral density.
In Section \ref{SmMT}, we consider kernel smoothing  the multi-taper
spectral estimate.
In Section \ref{SmLogMT}, the logarithm of the multi-taper
spectral estimate is kernel smoothed
to estimate the log-spectral density.
In Section \ref{ADAPT}, 
we consider a data adaptive variable bandwidth implementation 
of this method. 
In Section \ref{SIM}, we present our simulation results.
Sections \ref{REM} and \ref{SUM}
discuss and summarize our results.
In the appendix, we describe a new method for selecting the
initial halfwidth.

\section{STATISTICS OF MULTI-TAPER SPECTRAL ESTIMATORS}
\label{MTstat}

Every quadratic, modulation-invariant spectral estimator has the form
\begin{equation}\label{E1}
\hat{S}_{mt} (f) = \sum_{m,n=1}^N q_{mn} x_m x_n e^{2\pi i (m-n)f} \ ,
\end{equation} 
where ${\bf Q}=[q_{mn}]$ is a self-adjoint matrix
\cite{Bronetz,MullisScharf}.
Decomposing ${\bf Q}$ into its
eigenvector representation, ${\bf Q} = \sum_{k=1}^K \mu_k
\nubf^{(k)} \nubf^{(k)\dagger}$, 
(\ref{E1}) can be recast as
\begin{equation}\label{E2}
\hat{S}_{mt} (f) = \sum_{k=1}^K \mu_k \left| \sum_{n=1}^N \nu_n^{(k)}
x_n e^{-2 \pi inf} \right|^2 \ ,     
\end{equation}
where the {{$\nubf$}}$^{(k)}$ 
are the orthonormal eigenvectors of $\bf Q$
and  the $\mu_{k}$ are the eigenvalues. We call
(\ref{E2}) the multiple taper representation of the spectral estimate
\cite{PercWald,RST94,DJT82}.
(This name is often shortened to multi-taper and sometimes referred to
as a multiple spectral window estimate.)  
In practice,
quadratic spectral estimators are constructed by specifying the 
eigenvectors/tapers
and the weights. For concreteness, we will usually use the sinusoidal
tapers $\nu_m^{(k)} = \sqrt{{2 \over N+1}}
\sin \left({\pi k m \over N+1}\right) $ \cite{RiedSidMB}.
For these tapers, the spectral estimate (\ref{E2}) can be recast as
\begin{equation}\label{E3}
\hat{S}_{mt} (f) = \Delta \sum_{k=1}^K {\mu_k}
 |\zeta(f+k \Delta )-\zeta(f-k \Delta )|^2 \ ,         
\end{equation}
where $\Delta = \frac{1}{2N+2}$ and 
$\zeta(f)$ is the discrete Fourier transform
of $\{ x \}$: $\zeta(f) = \sum_{n=1}^N x_m e^{-2 \pi imf}$.
The corresponding smoothed periodogram estimate, $\hat{S}_{sp}(f) =
\sum_{k=-K}^K$ $|\zeta(f+k \Delta )|^2$ $ /(2KN+N)$, has an appreciably larger
bias. The sinusoidal multi-taper estimate reduces the bias since the
sidelobes of $\zeta(f+k \Delta )$ are partially cancelled by those of
$\zeta(f-k \Delta )$.

To analyze the multi-taper estimate, we use the local white noise 
approximation \cite{Koopmans}, 
which corresponds to assuming that the combined estimator of $\theta (f)$ has
its domain of dependence concentrated near frequency $f$. 
When $\mu_k = {1 / K}$,
$\Sh_{mt}(f)/S(f)$ has a $\chi^2_{2K}/(2K)$ distribution 
to leading order in $K/N$ \cite{ThomChave}.
 Note $
{\bf E  }\left[   \ln \left( \chi^2_{2K}/(2K) \right) \right] =
\psi (K)  - \ln(K)$,  
${\bf Var }\left[   \ln \left( \chi^2_{2K}/(2K) \right) \right] =\   \psi'(K)
$,  
where $\psi$ is the digamma function and $\psi'$ is the trigamma function.
The multi-taper estimate of the logarithm of the spectral density is
\begin{equation}\label{E3M}
\thh_{mt}(f) \equiv\ {\rm ln}[\Sh_{mt}(f)] - [\psi (K) -  \ln(K)] \ .
\end{equation} 


An alternative estimate of 
${\rm ln}[{S}^{{}} (f)]$ is to average 
the logarithms of the individual multi-taper estimates:
\begin{equation}\label{E4}
\overline{{\rm ln}[\hat{S}_{st}(f)]}  \equiv {1 \over K}
\sum_{k = 1}^{K}  {\rm ln}(
\frac{|\zeta(f+k\Delta) -\zeta(f-k\Delta)|^2}{2(N+1)} ) \ \ ,
\end{equation}     
where the subscript ``st'' denotes single taper.
Since the $\chi_2^2$ distribution has its most probable value
at zero, the distribution of its logarithm has a 
very long lower tail. This lower tail induces bias 
and increases the variance in the estimate: 
$ {\bf Bias [}  \overline{{\rm ln}(\hat{S}_{st})} {\bf ]} \simeq -0.577  
$, and \
${\bf Var [}  \overline{{\rm ln}(\hat{S}_{st})} {\bf ]} = 
\psi' (1)/K =   { \pi^2  / (6K)}$.        
By averaging the $K$ estimates prior to taking the
logarithm, we reduce both the bias and the variance. 
The variance reduction factor if one averages and then
takes logarithms, ${\ln[\Shb(f)]}$,  is $K\psi'(K)/\psi'(1)$. 
For large $K$, $K\psi'(K) \simeq 1 + \frac{1}{2K}$, so the variance
reduction factor (of reversing the order of the operations in 
(\ref{E4}) )
is asymptotically $6/\pi^2$.

The local bias of the multi-taper estimate is
\begin{equation}\label{E5}
{\bf E}[ \hat{S}_{mt} (f) -S(f)]\ \simeq {S^{\prime\prime} (f) \over 2}
 \sum_{k=1}^K \mu_k \int_{-\frac{1}{2}}^{\frac{1}{2}} |f^{\prime} |^2
| V^{(k)} (f^{\prime})|^2 df' \ ,    
\end{equation}
where the $k$-th spectral window, $V^{(k)}$, is the Fourier
transform of the $k$-th taper, 
$\nubf^{(k)}$: $V^{(k)} (f) =
\sum_{n=1}^N \nu_n^{(k)} e^{-2 \pi inf}$. Equation (\ref{E5}) neglects the
nonlocal bias and assumes $\sum_{k=1}^K \mu_k =1$. For the sinusoidal
tapers  with uniform weighting ($\mu_k =1/K$), (\ref{E5}) reduces to 
\begin{equation}\label{E6}
{\bf Bias} [ \hat{S}_{mt} (f) ]\ \underline{\sim}\
{S'' (f) \over 8} \sum_{k=1}^K \mu_k {k^2 \over N^2} \
=\ S''(f) {K^2 \over 24N^2}
\ ,            
\end{equation}
where the intermediate equality is derived in \cite{RiedSidMB}.
Noting that $S''(f)/S(f) = [\theta''(f) + |\theta^{\prime} (f)|^2]$,
the local bias of the estimate (\ref{E3M}) for the uniformly weighted 
sinusoidal tapers is
\begin{equation}\label{E65}
\Ebf[\hat{\theta}_{mt} (f) - \theta (f)]\ \underline{\sim}\ 
[\theta''(f) + |\theta^{\prime} (f)|^2] {K^2 \over 24N^2}
\ . \end{equation}

\section{SMOOTHED MULTI-TAPER ESTIMATE}
\label{SmMT}

We now consider kernel estimators of $\partial_f^{q}S(f)$
which smooth the multi-taper estimate: 
\begin{equation}\label{E7}
\widehat{\partial_f^{q}S_{\kap}} (f) \equiv \ {1\over h^{q+1}}
\int_{-\frac{1}{2}}^{\frac{1}{2}} \kappa \left( {f^{\prime} -f
\over h} \right) \hat{S}_{mt} (f^{\prime} ) df' \ ,      
\end{equation}
where the $\widehat{ \ }$ over ${\partial_f^{q}S_{\kappa}}$ denotes the
estimate of the $q$th derivative. The subscript on $\hat{S}_{\kappa}$
denotes the two-stage estimator constructed by first multi-tapering and then
kernel smoothing.
Here $\kappa (f)$ is a  kernel with Lipshitz  smoothness of degree 2 
with support in $[-1,1]$, and
$\kappa ( \pm 1)=0$.  The bandwidth parameter is $h$.
We say a kernel 
is of order $(q,p)$ if
$\int f^m \kap(f)df =  { m!\ }\delta_{m,q} \ , \ m= 0, \ldots,p-1$.
We denote the $p$th moment of a kernel of order $(q,p)$ by $B_{p} \equiv
\int f^p \kap(f)df /p!$. 
For function estimation ($q=0$), we  use $p=2$ and $p=4$. 
To estimate the second derivative,
we use a kernel of order (2,4).

Smoothing the multi-taper estimate replaces the original quadratic estimator
in (\ref{E1}) 
by another quadratic estimator, $\tilde{\bf Q}$  with
$\tilde{Q}_{mn} = \hat{\kappa}_{m-n} 
\sum_{k=1}^K \mu_k \nu_m^{(k)}\nu_n^{(k)}$,
where $\hat{\kappa}_m$ is the Fourier transform of the kernel smoother:  
$\hat{\kappa}_m \equiv h^{-(q+1)}\int \kappa ({f' \over h})e^{imf'} df'$. 
By Theorem 5.2 of Riedel \& Sidorenko \cite{RiedSidMB},  
this smoothed multi-taper estimator cannot outperform the pure multi-taper
method with minimum bias tapers.


\begin{thm} \label{thm1}
Let $S(f)$ be twice continuously differentiable with 
$0 < S_{min} \le S(f) \le S_{max} <\infty$. 
Consider the kernel smoothed multi-tapered spectral estimate (\ref{E7})
with $K$ tapers.
Let the kernel, $\kappa (f)$, be of order $(q,p)$ and have  
Lipshitz  smoothness of degree 2. 
Let the envelope of the spectral windows, $V^{(k)}(f)$, 
decay as $(Nf)^{-1}$ or faster for $f> K/N$ and assume that 
$\nu_{n+m}^{(k)} {\simeq}\ \nu_n^{(k)} [1+\ {\cal O}({Km \over N})]$.
Consider the limit that $N \rightarrow \infty$, $h\rightarrow 0$ and 
$K \rightarrow \infty$, such that ${K / (Nh)} \rightarrow 0$.
The kernel smoothed multi-tapered estimate (\ref{E7}) has
asymptotic variance:
\begin{equation}\label{Ethm1}
{\bf Var} \left[ \widehat{\partial_f^qS_{\kappa}} (f)\right] \simeq\ 
{\|\kappa\|^2 S(f)^2\over h^{2q+1} }
\sum_{k,k'=1}^K \mu_k \mu_{k'} \left( \sum_{n=1}^N
| \nu_n^{(k)}|^2 | \nu_n^{(k')} |^2 \right) \ 
+ \ {\cal O_R}\left(({K \over Nh})^{4/5} +\ ({h \over K})^{2}\right),     
\end{equation}
where
$\|\kappa\|^2 \equiv \int_{-1}^{1}\kappa(f)^2df$.
\end{thm}

We use the notation ${\cal O_R}(\cdot)$ to denote a  size of ${\cal O}(\cdot)$
relative to the main term. The condition, $\frac{K}{N}/h \rightarrow 0$,
implies that the smoothing from multi-tapering is much less than the smoothing
from kernel averaging. The condition that 
$\nu_{n+m}^{(k)} {\simeq}\ \nu_n^{(k)} [1+\ {\cal O}({Km \over N})]$
is fulfilled when the $k$-th
taper has a scale length of variation of $N/k$.
The sinusoidal tapers satisfy this condition as do the Slepian tapers
when their bandwidth parameter, $W$, is chosen as $K/N$.

\noindent{\em Proof:} 
We separate the variance into a broad-banded contribution 
$\approx 1/(N|f-f'|)^2$
for $|f -f'|>> K/Nh$ and a local contribution $\approx |f-f'|^2$.
The broad-band contribution is ${\cal O_R}(({hN \over KN})^{2})$.
The local contribution differs from a locally white process by
${\cal O_R}(S''(f)^{2}({K \over 2Nh})^2)$.  
We now consider the local contribution 
in the locally white noise approximation \cite{Koopmans}.
Using the Gaussian fourth moment identity and resumming yields
\begin{equation}\label{E8}
{\bf Var} \left[ \widehat{\partial_f^qS_{\kappa}} (f)\right] \simeq\ 
S(f)^2 {\rm tr} [\tilde{\bf Q}\tilde{\bf Q}]
\ =\ S(f)^2 \sum_{k,k'=1}^K \mu_k\mu_{k'} \sum_{n=1}^N \sum_{m=1-n}^{N-n} 
\hat{\kappa}_m^2\nu_{n+m}^{(k)} \nu_n^{(k)} \nu_{n+m}^{(k')} \nu_n^{(k')} \ .
\end{equation}          
Our kernel$, \kap(\cdot)$ is Lipshitz of degree 2, and therefore
 $\hat{\kappa}_m \sim\ {\cal O}(\|\hat{\kap} \| /(mh)^2 )$ for $mh \gg 1$.
Expanding $\nu_{n+m}^{(k)}$ in $mK/N$ and truncating in $m$  yields
$$
{\bf Var} \left[\fqShk(f)\right]\       
\underline{\sim}\ S(f)^2 \sum_{k,k'=1}^K \mu_k \mu_{k'} \left( \sum_{n=1}^N
| \nu_n^{(k)}|^2 | \nu_n^{(k')} |^2 \right) \left(
\sum_{m=1}^N \hat{\kappa}_m^2 \right) \         
$$ 
\begin{equation}\label{E9}
=\ S(f)^2 {\|\kappa\|^2 \over h^{2q+1} }
\sum_{k,k'=1}^K \mu_k \mu_{k'} \left( \sum_{n=1}^N
| \nu_n^{(k)}|^2 | \nu_n^{(k')} |^2 \right) \ .       
\end{equation}
The first  line is valid to ${\cal O}(1/(mh)^4 ) + {\cal O}({Km / N})$, so we
Taylor expand $\nu_{n+m}^{(k)}$ for $|mh |<{\cal O}((Nh/K)^{1/5})$ and
drop all terms with $|mh|>{\cal O}((Nh/K)^{1/4})$. The resulting expression is
accurate to ${\cal O_R}((K/Nh)^{4/5})$. The final line follows from
Parseval's identity. \ \ \ \ $\Box$

For $K=1$, Eq.~(\ref{E9}) reduces to the well known result \cite{Walden90} 
for the variance of smoothed tapered periodogram:
\begin{equation}\label{E10}
{\bf Var} \left[ {1\over h^{q+1}}
\int \kappa ({f-f' \over h})|\zeta_{\nu} (f')|^2 df'\right]
\ \underline{\sim}\
{S(f)^2\|{\kappa}\|^2 \over h^{2q+1}} \sum_{n=1}^N \nu_n^4
\ ,          
\end{equation}
where $\zeta_{\nu} (f)$ is the tapered Fourier transform.
In (\ref{E10}), $ \sum_{n=1}^N \nu_n^4$ is ${\cal O}(1/N)$. 
For the sinusoidal tapers, (\ref{E9}) can be
explicitly evaluated: 
\begin{equation}\label{E12}
{\bf Var} \left[ \fqSh_{\kap}(f)\right]\ \underline{\sim}\
{\|{\kappa}\|^2S(f)^2 \over Nh^{(q+1)} }\left(1+ {1\over 2K}\right)
+ {\cal O_R}\left(({h \over K})^2\right)
+\ {\cal O_R}\left(({K \over Nh})^{4/5}\right)
\ ,          
\end{equation}
where we have used
\begin{equation}\label{E11}
{1 \over K^2}
\sum_{k,k'=1}^K  \sum_{n=1}^N
| \nu_n^{(k)}|^2 | \nu_n^{(k')} |^2 =
{4 \over K^2(N+1)^2} \sum_{k,k'=1}^K \sum_{n=1}^N 
\sin ( {\pi kn \over N+1})^2 \sin ({\pi k'n \over N+1})^2  
\ = {2K+1  \over  2K(N+1) }
\ .    \nn
\end{equation}

\section{SMOOTHED LOG MULTI-TAPER ESTIMATE}
\label{SmLogMT}

We now show that combining kernel smoothing with 
{\em multi-tapering does improve
the estimation of the  logarithm} of the spectral density, 
$\theta (f) = \ln[S(f)]$.  Let
\begin{equation}\label{E13}
\widehat{\partial_f^q\theta}_{\kappa} (f)
\ \equiv\ {1\over h^{q+1}} \int_{-\frac{1}{2}}^{\frac{1}{2}} \kappa \left( {f^{\prime} -f
\over h} \right) \hat{\theta}_{mt} (f^{\prime} ) df' \ .
\end{equation}
For $h \ll 1$, and $Nh \gg 1$, we expand (\ref{E13}) in the bandwidth
\begin{equation}\label{E14}
{\bf Bias} [ \widehat{\partial_f^q\theta}_{\kap} (f)] 
\simeq B_{p} \partial_f^p \theta(f) h^{p-q }  +
\partial_f^q[ \theta''(f) + | \theta^{\prime}(f) |^2 ] {K^2 \over 24N^2} \ .
\end{equation} 
The first term is the bias from kernel smoothing
and the second term is from the sinusoidal multi-taper estimate (\ref{E65}).
Traditionally, the ``delta approximation'',
${\bf Var}[f(X)] = f'({\bf E}[X])^2 {\bf Var}[X]$,
is used to evaluate the variance of the smoothed log-periodogram.
For the delta approximation to be valid,  
the characteristic scale of variation of $f(\cdot)$
must be large relative to $\sqrt{{\bf Var}[X]}$, 
where $f$ is continuously differentiable. This requirement
is not fulfilled for the log-periodogram, 
and the resulting  analysis makes an order one error 
in single taper estimation.
For the multi-taper estimation,
the expansion parameter for the delta approximation is $1/K$. 
To leading order in the $1/K$ expansion, the variance inflation factor from
the long tail of the $\ln[\chi^2_{2K}]$ distribution is not visible.
Recall that ${\bf Var} [ \hat{\th}_{mt} (f')]
\approx \  [K\psi'(K)] \times {\bf Var} [ \hat{S}_{mt} (f')]/ S(f)^2$
for $|f-f'| <<1$.
We believe that adding a $K\psi'(K)$ correction improves the
accuracy of the delta approximation
for $f'' \ne f'$.
Therefore, we evaluate the variance of the smoothed log multi-taper estimate
by using the approximate identity:
\begin{equation}\label{E153}
{\bf Cov} [ \hat{\th}_{mt} (f'),\hat{\th}_{mt} (f'')]
\approx \  {[K\psi'(K)] \over S(f)^2}
\times {\bf Cov} [ \hat{S}_{mt} (f'),\hat{S}_{mt} (f'')]
\ ,
\NEQ
for $|f'-f| <<1$ and $|f''-f| <<1$.
Using (\ref{E153}),
the variance of $\hat{\theta} (f)$ is
\begin{equation}\label{E15}
{\bf Var} [ \widehat{\partial_f^q\theta}_{\kappa} (f)]
\ \underline{\sim}\  {K\psi'(K)
\over S(f)^2h^{2(q+1)}} \int_{-\frac{1}{2}}^{\frac{1}{2}}
\int_{-\frac{1}{2}}^{\frac{1}{2}} \kappa \left( {f-f^{\prime} \over h} \right) \kappa 
\left( {f-f'' \over h} \right) {\bf Cov}[ \hat{S}_{mt} (f^{\prime} ),
\hat{S}_{mt}(f'' )] df'df''
\ . \end{equation}           
Thus, 
the variance of $\fqthh(f)$ reduces to
the same calculation as the variance of $\fqSh(f)$:
\begin{equation}\label{E152}
{\bf Var} \left[ \fqthh_{\kap}(f)\right]\ \underline{\sim}\
{(K + \frac{1}{2})\psi'(K) \|{\kappa}^2\| \over Nh^{2q+1} }
+ {\cal O_R}\left({1 \over K}\right)
+\ {\cal O_R}\left(({K \over Nh})^{4/5}\right)
\ ,          
\end{equation}
for the uniformly weighted sinusoidal tapers.
(See  the calculation in Theorem \ref{thm1}.)
Combining (\ref{E14}) with  (\ref{E15}) 
yields the expected asymptotic square error (EASE) in 
$\widehat{\partial_f^q \th}_{\kappa}$:

\begin{thm} \label{thm2}
Let $S(f)$ have $p$ continuous derivatives.
Consider the two-stage estimate (\ref{E13})
using the uniformly weighted sinusoidal tapers in the first-stage.
Under the hypotheses of Theorem \ref{thm1} and the formal approximation
(\ref{E153}), 
the expected asymptotic square error 
of $\widehat{\partial_f^q \th}_{\kappa}$   is  
\begin{eqnarray}
\label{E17}
{\bf E}\left[\left|\widehat{\partial_f^q \th}_{\kap}(f) -
{\partial_f^q \th}(f) \right|^2\right] \ \approx \
\left[ B_{p} \partial_f^p \theta(f) h^{p-q }  +
\partial_f^q[ \theta''(f) + | \theta^{\prime}(f) |^2 ] {K^2 \over 24N^2}
\right]^2
\nn \\
\ + \
{(K + \frac{1}{2})\psi'(K)
\|\kap\|^2 \over Nh^{2q+1}} 
\ +\ {\cal O_R}\left(h^{2(p-q)+1}\right) 
+ {\cal O_R}\left({1 \over K}\right)+\ 
{\cal O_R}\left(({K \over Nh})^{4/5}\right) 
\ . 
\end{eqnarray}
\end{thm}
The benefit of multi-tapering (in terms of the  
variance reduction) is significant for using a 2 to 20 tapers. However,
the marginal benefit of each additional taper    
tends rapidly to zero. 
Minimizing (\ref{E17}) with respect to $h$ and $K$ yields
the following result:

\begin{cor}
Under the hypotheses of Theorem \ref{thm2},
the expected asymptotic square error (EASE)
of $\widehat{\partial_f^q \th}_{\kappa}$  is minimized by  
\begin{equation}\label{E18} 
h_{o}(f) = \left[ {2q+1 \over 2(p-q)}
{(K + \frac{1}{2})\psi'(K) \|\kap\|^2 \over  
B_{p}^2 N |\partial_f^p \th(f)|^2 }
\right]^{1\over 2p+1}
\ , \end{equation}           
and
\begin{equation}\label{E19}
B_{p} [ \partial_f^p \theta(f) ]
\{ \partial_f^q[ \theta''(f) + | \theta^{\prime}(f) |^2 ]\} \ 
K_{opt}^3 \ \simeq\ 6 \|\kap\|^2  N h_o^{-(p+q+1)}
\ .       
\end{equation}
\end{cor}

Thus $h_{opt} \sim N^{-1/(2p+1)}$ and  $K_{opt} \sim N^{(3p+q +2)/(6p+3)}$.
For kernels of order $(0,2)$, this reduces to
$h_{opt} \sim N^{-1/5}$ and  $K_{opt} \sim N^{8/15}$.
Thus the ordering $1 \ll K \ll Nh$ is justified.
The EASE (\ref{E17}) depends only weakly on $K$ for
$1 \ll K \ll Nh$ while the dependence on the choice of bandwidth is strong.
When the bandwidth, $h_o$, satisfies (\ref{E18}),
the leading order EASE  reduces to 
\begin{equation}\label{E21}
{\bf E}\left[\left|\widehat{\partial_f^q \th}(f_j) -
{\partial_f^q \th}(f_j) \right|^2\right] \ \simeq \
M_{q,p}  |B_{p}\partial_f^p \th(f_j)|^{2(2q+1)\over(2p+1)}
\left({  (K + \frac{1}{2})\psi'(K) \|\kap\|^2 \over N } 
 \right)^{2(p-q)\over(2p+1)}
\ , \end{equation}           
where $M_{q,p}\equiv ( {2q+1 \over 2(p-q)} )^{2(p-q)\over(2p+1)} +
( {2(p-q) \over 2q+1} )^{(2q+1)\over(2p+1)}$.
Thus the EASE in estimating 
$\partial_f^q \th$ is proportional to $N^{-2(p-q)\over(2p+1)}$.
We note that if $K=1$ (a single
taper), the variance term in (\ref{E17}) is inflated by a factor of
${\pi^2 \over 6} \sum_{n=1}^N \nu_n^4$. Thus using a moderate level of
multi-tapering prior to smoothing the logarithm reduces the EASE by a
factor of $[{\pi^2 \over 6} \sum_{n=1}^N \nu_n^4 ]^{4/5}\
=\ [{\pi^2 \over 4}]^{4/5}$, where we substitute
$ \sum_{n=1}^N \nu_n^4 =\ 1.5 $ for the sinusoidal tapers.

From (\ref{E21}), using the best fixed halfwidth kernel smoother 
degrades performance by a factor of 
\begin{equation}\label{E22} 
{ {\rm EASE}(h_{global}) \over {\rm EASE}(h_{variable})} \ = \
\left[\int_{-\frac{1}{2}}^{\frac{1}{2}}|\theta^{\prime\prime}(f)
|^2 df\right]^{1/5}\left/ \int_{-\frac{1}{2}}^{\frac{1}{2}}
|\theta^{\prime\prime}(f)|^{2/5}df \right. 
\end{equation}         
over using an optimal variable halfwidth smoother \cite{MS87}.
In many cases, the spectral
range is large, and thus it is often essential to allow the bandwidth 
to vary locally as a function of frequency.

Equation (\ref{E18}) gives an explicit solution for the bandwidth which
minimizes the local bias versus variance trade-off.
It shows that when $\th(f)$ is rapidly varying ($|\th''(f)|$ is large), 
then the kernel bandwidth should be decreased. 
However, 
(\ref{E18}) has  two major difficulties.
First, (\ref{E17})-(\ref{E21}) are based on a Taylor series
expansion and the expansion parameter is $h_{o} \sim 1/N^{1/(2p+1)}$.
Even when $1/N$ is small, 
$(1/N)^{1/(2p+1)}$ may be not so small. Second, 
$\th''(f)$ and $h_{o}(f)$ are unknown and need to be estimated.

\section{DATA-ADAPTIVE ESTIMATE} 
\label{ADAPT}

In practice, $\theta^{\prime\prime}(f)$ is unknown and we use a data-adaptive
multiple stage kernel estimator where a pilot estimate of the optimal
bandwidth is made prior to estimating $\theta (f)$. 
To simplify the implementation, we choose $K$ 
independent of frequency, and usually set $K\approx \ c N^{8/15}$,
where $c$ is a constant.
For nonparametric function estimation, data adaptive multiple stage schemes 
are given in \cite{Brockmann,MS87,RiedSidCIP}.
A straightforward application of these schemes 
to multi-taper spectral estimation has the following steps:


0)  Evaluate the multi-taper estimate of (\ref{E3}) on a grid of size $2N+2$.
If the computational effort is not important, set $K=N^{8/15}$; otherwise
choose $K$ according to your computational budget.

1a) Kernel smooth $\hat{\theta}_{mt} (f)$ with a kernel of order (0,4) for
a number of different bandwidths, $h_{\ell}$, and evaluate the average
square residual (ASR) as  a function of $h_{\ell}$:
\begin{equation}\label{E23}
ASR(h_{\ell})=\sum_{n=1}^N 
|\hat{\theta}_{st}(f_n )-\hat{\theta}_{\kap}(f_n |h_{\ell})|^2 \ ,
\end{equation}
where $\hat{\theta}_{\kap}(f_n |h_{\ell})$ 
is the kernel estimate of $\theta (f)$
using bandwidth $h_{\ell}$ applied to $\hat{\theta}_{mt}(f)$,
while $\hat{\theta}_{st}(f_n )$ is the single taper estimate:
$\hat{\theta}_{st} (f)=\ln [|\zeta(f+ \Delta )-\zeta(f-\Delta )|^2 /2(N+1)]\
+.577$.

1b)
Estimate the optimal (0,4) global halfwidth using a goodness of fit
method. Relate this to the optimal (2,4) using the halfwidth quotient relation.
(See below.)

2) Estimate $\theta''(f)$ by  smoothing the multi-taper estimate
with global halfwidth $h_{2,4}$.

3) Estimate $\theta (f)$ by substituting $\hat{\theta}^{\prime\prime} (f)$
into the optimal halfwidth expression corresponding to the minimum of (14).

For Step 1b), M\"uller and Stadtm\"{u}ller propose to determine 
the starting halfwidth by minimizing the Rice criterion. 
In \cite{RiedSidCIP}, 
we describe a different method for selecting the initial bandwidth in step 1b).
Our method is based on fitting the average square residual of (\ref{E23})
to a parametric expression based on (\ref{E17}).  This parametric fit
usually outperforms the Rice criterion because it uses an asymptotically
valid expression.

In (\ref{E23}), the ASR is computed relative to the single taper
estimate, $\hat{\theta}_{st}(f_n )$, instead of the multi-taper
estimate, $\hat{\theta}_{mt}(f_n )$. We do this because the
multi-taper estimate is strongly autocorrelated for frequencies,
$f$ and $f'$ with $|f-f'|\le K/2N$.
To correct for using $\hat{\theta}_{st} (f)$ in step 1 and 
$\hat{\theta}_{mt} (f)$ in steps 2 and 3 , we 
inflate the variance in the (0,4) kernel estimate. 
The halfwidth quotient relation 
relates the optimal halfwidth for derivative estimates,
$\hh_{2,4}$ to the optimal halfwidth for a $(0,4)$ kernel using (\ref{E18}):
\begin{equation}\label{E24}
 \hh_{2,4} = H(\kap_{2,4},\kap_{0,4}) \hat{h}_{0,4} , \ \ 
{\rm where} \  H(\kap_{2,4},\kap_{0,4}) \equiv 
\left({ 10 B_{0,4}^2 \|\kap_{2,4}\|^2\over
B_{2,4}^2 \|\kap_{0,4}\|^2 }\right)^{1\over 9}
\left({ \pi^2 N \sum_n |\nu^{(1)}_n|^4\over
6 } \right)^{1\over 9}
\ . \end{equation}      
The last term in parentheses 
 is the variance inflation factor from using a single taper. 
To minimize the effects of tapering-induced autocorrelation, we
recommend using a Tukey taper for $\hat{\theta}_{st}$.

 When $\hat{\theta}^{\prime\prime}(f)$ is vanishingly small, the optimal
halfwidth becomes large. Thus, $\hh_{0,2}$ needs to be regularized. 
Following \cite{RiedSidCIP}, 
we determine the size of the
regularization from $\hat{h}_{0,4}$ in the previous stage.

We say a ``plug-in'' scheme has a relative convergence rate 
of $N^{- \alpha}$ if
$${\bf E}\left[|\hat{\th}(f|\hh_{0,2}) - \th(f)|^2 \right] \
\simeq \ \left( 1+ {\cal O}(C_r^2 N^{-2\alpha})\right)
{\bf E}\left[|\hat{\th}(f|h_{0,2}) - \th(f)|^2\right]
\  , 
$$
where $h_{0,2}$ is the optimal halfwidth and $\hat{h}_{0,2}$ is the
estimated halfwidth.
In \cite{Brockmann}, 
a detailed analysis of the convergence properties of their
similar scheme is given. 
Their scheme has an optimal convergence rate of $N^{-4/5}$
and a relative convergence rate of $N^{-1/4}$. Our  simpler method
has the same convergence rate of $N^{-4/5}$ and
a slightly slower relative convergence rate: $N^{-2/9}$.

\section{COMPARISON OF KERNEL SMOOTHER ESTIMATES}
\label{SIM}

We now compare three different kernel smoother estimates
of the log-spectrum: 
1) Kernel smoothing the log-multitaper estimate, $\thh_{mt}$
as in (\ref{E13}); 
2) Kernel smoothing the log-single taper estimate, $\ln[\Sh_{st}]$; 
3) The logarithm of the kernel smoothed multi-taper spectral estimate, 
$\ln[\Sh_{\kappa}]$ as in (\ref{E17}). 
In all cases, we use a variable halfwidth kernel smoother with
the initial $h_{0,4}$ halfwidth estimated by the fitted square residual
method as described in Sec.~\ref{ADAPT} and the appendix.

We use the 
moving average time series model which was
considered in \cite{OSullivan}: 
$x_t = e_t - 0.3e_{t-1}  -0.6e_{t-2} +\ 0.3e_{t-3} $,
where 
$e_t$ is a zero mean, unit variance, uncorrelated Gaussian process.
We compute the integrated square error (ISE): $\int |\thh(f) -\th(f)|^2df$,
averaged over 500 realizations 
for time series lengths of 128 and 1024.
We use the sinusoidal tapers and choose $K = (N/2)^{8/15}$, which is $K=9$ for
$N=128$ and $K=28$ for $N=1024$.
Table 1 summarizes our simulation:

\bigskip

\begin{tabular}{|c|c|c|c|c|}
\hline
Error Criterion & MISE & MaxISE  & MISE & MaxISE  \\
\hline
 Method & $N=128$ & $N=128$ & $N=1024$ & $N=1024$ \\
\hline
Smoothed log-multi-taper (\ref{E13})   & .453 & .694 & .186 & .515  \\
Log of smoothed multi-taper (\ref{E7}) & .483 & .743 & .195 & .515  \\
Smoothed log-single taper              & .622 &1.009 & .209 & .842  \\
\hline
\end{tabular}

\bigskip

Table 1: Integrated square error averaged over 500 realizations
where MaxISE is the integrated square error for the worst of the
500 realizations.

\medskip

The simulation shows that smoothing  before taking the logarithm of
the multitaper estimate performs somewhat more poorly than smoothing
the log multi-tapered estimate. The performance degradation is
6.6 \% for $N=128$ and 4.8 \% for $N=1024$.
The performance differential is due to the presence of broad-band bias
error. As $N$ increases, the smoothing halfwidth decreases and the
effects of  broad-band bias will shrink. For more peaked spectral densities,
$N$ may have to be quite large before the two estimates perform similarly.

In comparing the first and third estimates, we expect to
see an improvement factor of $[{\pi^2 \over 4} ]^{.8}$ for
multi-tapering.
Multi-tapering prior to smoothing the logarithm
reduces the ISE by more than  expected. 
We attribute this additional reduction to
the poor performance of automatic halfwidth selection criteria
in the presence of strong noise. 
Note that using a single taper is very nonrobust in the sense that
the worst realizations have much larger ISEs than do either of the 
other two methods. 
Our simulations also indicate that the optimal number of tapers grows at
faster than $N^{8/15}$ for our particular spectrum and $100 < N < 1000$.

\section{REMARKS} \label{REM}



1) P.~Bloomfeld (private correspondence) points out that a  similar
variance reduction can be achieved  by pre-smoothing the periodogram
before transforming to the logarithmic scale and smoothing again. Our
analysis in Sec.~III shows the optimal amount of pre-smoothing. Note
multi-tapering offers broad-band bias protection with asymptotically
no variance inflation. In contrast, pre-smoothing the tapered periodogram
inflates the variance by $\sum_n |\nu^{}_n|^4$. For the pre-smoothing
algorithm to be as efficient as multi-tapering, the amount of tapering
needs to go to zero as $N \rightarrow \infty$. 
												
2) Pawitan and O'Sullivan   \cite{OSullivan} 
advocate a penalized Whittle likelihood
estimate with generalized cross-validation. 
Clearly, it should be advantageous to use an approximation
of the likelihood. Unfortunately, the penalized likelihood approach
corresponds to a $fixed$ halfwidth kernel and
does not reduce the strength of the smoothing near the points of rapid spectral
variation. We expect a variable halfwidth kernel smoother to outperform
a penalized likelihood method by the factor given in (\ref{E22}). 
Also note that the Whittle likelihood is asymptotic and provides
with no information on the amount of tapering which should be done in
a finite sample size.

3) An early adaptive multi-taper scheme was proposed in \cite{DJT82}.
This scheme makes the unrealistic assumption 
that the spectral density is $S(f)$
in the region $[f-W,f+W]$ and is $(\sigma^2 -2W S(f))/(1-2W)$
elsewhere, where $W$ is a bandwidth parameter and $\sigma^2$ is
the variance. 
Furthermore, the adaptive weighting of
\cite{DJT82} is usually computed with the Slepian tapers,
which have a fixed bandwidth, $W$. The goal of adaptive methods, to
reduce the bandwidth of the estimate when the spectrum is rapidly varying,
is defeated by the inflexibility of the Slepian tapers. 
In our previous simulations \cite{RST94,RiedSidMB},
the adaptive weighting of \cite{DJT82} has performed so poorly
that we no longer consider it a viable alternative.

4) The evolutionary spectrum of Priestley \cite{Priestley65} 
can be estimated by
applying a two dimensional kernel smoother (in the time-frequency plane)
to the log-multi-tapered spectrogram (Riedel \cite{RiedelEvSp}).
 

\section{SUMMARY} \label{SUM}

We have analyzed the expected asymptotic square error of the smoothed
log multi-tapered periodogram and shown that multi-tapering reduces the
error by a factor of $[{\pi^2 \over 4}]^{4\over 5}$ for the sinusoidal tapers.
The optimal rate of pre-smoothing prior to taking logarithms is $K \sim
N^{8/15}$, but the expected loss depends only weakly on $K$ when
$1 \ll K \ll Nh$. A similar enhancement in performance has been
reported by Walden \cite{Walden95} for estimating the innovations variance:
$\exp\left[\int \ln[S(f)]df \right]$.

We have proposed a data-adaptive multiple stage variable halfwidth kernel
smoother. It has a relative convergence of $N^{-2/9}$, which can be improved
to $N^{-1/4}$ if desired by using the iteration method of  [2].
Our multiple stage estimate has the following steps:
1) Estimate the optimal kernel halfwidth for a kernel of (0,4) 
for the log-single tapered periodogram.
2) Estimate $\hat{\theta}_{mt}(f)
\equiv \ln [\hat{S}_{mt}(f)]-B_K $ as described in Sec.~2.  
3) Estimate $\theta^{\prime\prime}(f)$ using a kernel smoother of order
(2,4). 4) Estimate $\theta (f)$ using a kernel smoother of order (0,2) with
the halfwidth 
$h_0 (f)\approx c| \widehat{\partial^2_f \theta} |^{-2/5}N^{-1/5}$. 



\

\appendix{\bf APPENDIX: FITTED SQUARE RESIDUAL INITIALIZATION}

The factor method (\ref{E24}) relates the optimal halfwidth for a
(2,4) kernel to that of a (0,4) kernel.
To begin the kernel estimation, a halfwidth for the (0,4) kernel
needs to be specified. 
In \cite{MS87}, M\"uller and Stadtm\"{u}ller propose to select $h_{0,4}$
using a 
penalized goodness of fit (GoF) method such as
generalized cross-validation or the Rice criterion.
In penalized goodness of fit methods, the (0,4) halfwidth is
chosen by minimizing a functional of $Nh$ and  $ASR(h)$ (\ref{E23}). 


Unfortunately, 
these GoF functionals are often  flat near their minimum
and the actual minimum can be
very sensitive to noise. As a result, the halfwidth given by the
GoF methods tends to vary appreciably even when the noise is weak.
Furthermore, when tapering or multitapering is used, the residual
errors are correlated and GoF methods have great difficulty estimating 
the optimal halfwidth.
To  remedy this sensitivity problem, 
we fit  $ASR(h)$ to a two parameter model prior to
estimating the optimal bandwidth \cite{RiedSidCIP}. 

The fitted residual error method \cite{RiedSidCIP} begins by
evaluating the average square residual  (ASR) (\ref{E23}) 
as a function of the kernel halfwidth. (GoF methods also evaluate
$ASR(h)$.)
For the $(0,4)$ kernel, the bias error is proportional to $h^{4}$
our parametric model is
$$
ASR (h) \; \sim \; a V(h) + b h^{8}
 \; ,
\eqno(A1)$$
where
$V(h) = \sum_{j=1}^N (\mu_j(h)-\delta_{0,j})^2 $ with
$\mu_j(h) = \kap(j/Nh) /h$. In the large $Nh$ limit,
$V(h)\approx
 1 +  [\|\kappa\|^2 -2\kappa (0)]/Nh$.
Equation (A1) represents the integral of (\ref{E17}) over  frequency.
The first term corresponds to the bias, $\int |B_{p}\fpth|^2df$, 
and the second term corresponds to the variance.
The model has two parameters, $a$ and $b$. (Note that for smoothing
the log-tapered periodogram of a Gaussian time series, $a= 1$.)  

By parameterizing $ASR (h)$ with (A1), 
we are assured of an unique minimum.
The variance of $ASR (h)$ is of order $\frac{1}{N}$
and is practically independent of $h$.
We determine $a,b$ by minimizing the weighted least squares problem:
$$\{ a,b\}  =
{\rm argmin}_{\{ a,b\} } \sum_{h_j}
  \left[ ASR(h_j) - \left( a V(h_j) + b h_j^{8} \right) \right]^2
\ ,\eqno(A2)$$
where we use an equi-spaced grid in $h$. The upper and lower limiting
bandwidths, $h_U$ and $h_L$,
for the grid in $h$ is chosen such that
$ASR(h_U)\approx 2ASR(h_{min}) \approx ASR(h_L)$.
The least squares fit in (A2) is heuristic because the residual error
are correlated for different values of $h$. 

The $ASR$ measures the difference between the measured values and the
prediction based on the same  measured values.
We wish to minimize the difference between the predicted values and
new measurements. The expected value of the $ASR$ differs from the 
EASE (\ref{E17}) by a function of $Nh$.
We then choose the halfwidth which minimizes our parameterized model
of the EASE: 
$h_{opt} = \left({a\|\kap\|^2}/{8b}\right)^{1/9}$.
We caution that the theoretical convergence properties
of this estimator are unknown.
Nevertheless, our simulations show
that this fitting procedure gives more stable halfwidth
estimates than penalized goodness of fit methods do.
The advantage of the fitted residual error method appears even larger
when the residuals are correlated from tapering.



\begin{thebibliography}{99}





\bibitem{Brockmann}{T.~Brockman,  Th.~Gasser and E.~Hermann, 
 {``Locally adaptive bandwidth choice for kernel regression estimators,''}
{\em J. Amer. Stat. Assoc.}, {vol.~88}, pp.~1302-1309, Dec.~1993.}

 \bibitem{Bronetz}    T.~P.~Bronez, 
{\em Nonparametric Spectral Estimation of Irregularly Sampled
Multidimensional Random Processes}, PhD Thesis,
Arizona State University, 1985.












\bibitem{Koopmans}{L.~H.~Koopmans, 
{\em The spectral analysis of time series.} Ch.~8, 
{New York: Academic Press},  1974.  }





\bibitem{MS87}{H~.G.~M\"uller and U.~Stadtm\"uller, 
 {``Variable bandwidth kernel estimators of regression curves,"}
{\em Annals of Statistics}, {vol.~ 15}, pp.~182-201, Jan.~1987.} 


\bibitem{MullisScharf}{C.~T.~Mullis  and  L.~L.~Scharf,  
{``Quadratic estimators of the power spectrum,''}
in {\em Advances in spectrum analysis},
edited by S.~Haykin, New York: Prentice-Hall, 1991, 
Ch.~1, {pp.~1-57}.}  






\bibitem{OSullivan}{Y.~Pawitan and F.~O'Sullivan, 
``Penalized Whittle likelihood estimation of spectral density functions,''
{\em J.~Amer.~Stat.~Assoc.}, {vol.~89}, pp.~600-610, June 1994.}


\bibitem{PercWald}{D.~Percival and A.~Walden, 
{\em Spectral Analysis for Physical Applications:
Multi-Taper and Conventional Univariate Techniques}.
Cambridge: Cambridge University Press, 1993. }



\bibitem{Priestley65} 
{M.~B.~Priestley, 
{ Evolutionary spectra and nonstationary processes.} 
{\em J. Roy. Stat. Soc. Ser. B}, {vol.~27}, pp.~204-237, March 1965.} 


\bibitem{RiedelEvSp} 
{K.~S.~Riedel, 
{``Optimal kernel estimation of evolutionary spectra,"} 
{\em I.E.E.E.~Trans.~on Signal Processing}, {vol.~41}, pp.~2439-2447,
July 1993.} 

\bibitem{RST94}  K.~S.~Riedel, A.~Sidorenko, 
D.~J.~Thomson,
``Spectral  density estimation of plasma 
fluctuations I: Comparison of Methods,'' 
{\em Physics of Plasmas}, {vol.~1}, pp.~485-500, March 1994.



\bibitem{RiedSidCIP}{K.~S.~Riedel and A.~Sidorenko, 
``Data adaptive kernel smoothers- how much smoothing?''
{\em Computers in Physics}, 
{vol.~8}, pp.~402-409, May 1994. }

\bibitem{RiedSidMB}{K.~S.~Riedel and A.~Sidorenko, 
``Minimum bias multiple taper spectral estimation,''
{\em I.E.E.E.~Trans.~on Signal Processing}, {vol.~43}, pp.~188-195, Jan.~1995.}







\bibitem{DJT82}{D.J.~Thomson, 
{``Spectrum estimation and harmonic analysis,''}
{\em  Proc.~I.E.E.E.}, {vol.~70}, pp.~1055-1096, Sept.~1982. }  


\bibitem{ThomChave}
{D.~J.~Thomson and A.~D.~Chave, 
{``Jackknife error estimates for spectra, coherences 
and transfer functions,''} 
in {\em Advances in spectrum analysis},
edited by S.~Haykin, New York: Prentice-Hall, 1991, 
Ch.~2,  pp.~58-113.} 







\bibitem{Walden90}{A.~Walden,      
``Variance and degrees of freedom of a spectral estimator
following data tapering and spectral smoothing,''
{\em Signal Processing}, 
{vol.~20}, pp.~67-75, May 1990.} 


\bibitem{Walden95}{A.~Walden,     
``Multitaper estimation of the innovation  variance of a stationary
time series,''
{\em I.E.E.E.~Trans.~on Signal Processing}. 
{vol.~43}, pp.~181-187, Jan.~1995.}

\end{thebibliography}
\end{document}